# Implementation of SNR estimation based Energy Detection on USRP and GNU Radio for Cognitive Radio Networks

Jonti Talukdar, Bhavana Mehta, Kinjal Aggrawal, Mansi Kamani
Department of Electronics and Communication Engineering,
Institute of Technology, Nirma University, Ahmedabad, India
Email: {14bec057, 14bec028, 13bec054, 13bec057}@nirmauni.ac.in

*Abstract*—Development of smart spectrum sensing techniques is the most important task in the design of a cognitive radio system which uses the available spectrum efficiently. The adaptive SNR estimation based energy detection technique has the dual benefit of improving the efficiency of spectrum usage by capitalizing on the underutilization of the spectrum in an adaptive and iterative fashion, as well as reducing the hardware resources leading to easy implementation on a versatile and diverse group of cognitive radio infrastructures. The use of adaptive threshold for energy detection based on SNR estimation improves the spectrum sensing performance and efficiency of the cognitive radio by many folds, especially in low SNR as well as high noise variance situations. The proposed method is implemented on the USRP B200 and results show significant improvement in the detection rate of primary users as compared to conventional energy detection techniques.

## I. INTRODUCTION

The fixed spectrum policy which initially started out as an ideal solution to manage multiple users over a wide spectral bandwidth has in reality become the reason for its inefficient use. Largescale establishment and adoption of several concurrent communication channels has created the problem of network congestion thereby reducing the quality of service to end users. Exponential growth in communication infrastructure has also created a need for higher data rates. This has led to the dual problem of heavy congestion in certain frequency bands as well as underutilization in other bands thus leading to spectrum scarcity [1].

For static allocations, the overall spectrum utilization varies from 15% to 85% (as per Federal Communications Commission) and remains idle for most of the time thus showing considerable scope for improvement [2]. The overall framework of Dynamic Spectrum Access (DSA) techniques allows for the opportunistic use of the unutilized part of the spectrum by unlicensed users ensuring minimal interference and improving the overall spectrum efficiency. Cognitive radio (CR) acts as the key technology enabler for the implementation of DSA techniques through real time interaction with the active radio environment [3]. This includes identification of unused parts of the spectrum, a technique known as spectrum sensing. It relies on the search and identification of spectrum holes, i.e. unoccupied areas of the spectrum in both spatial as well as temporal domains. The cognitive radio, upon detecting the absence of a licensed user within a frequency band allocates that vacant band to secondary unlicensed users without causing any interference to the primary licensed users, thus ensuring optimal quality of service for all users as well as maximizing spectrum utilization. A cognitive radio's ultimate objective thus, is to deliver the best utility of the available spectral resources in terms of both spectral efficiency as well as flexibility. In order to achieve this objective, it is imperative to use a smart spectrum sensing technique which is both robust in terms of accuracy of detection as well as computationally inexpensive to facilitate easy implementation.

The development of smart communication systems necessitates the use of a mixed approach with secondary users having cognitive radio capabilities for real time analysis as well as the efficient management of the available spectral bandwidth, which is a limited resource. At the same time, it is also essential to ensure that the CR framework operates under a lower infrastructure requirement in order to facilitate wide scale adoption and portability [4]. Several spectrum sensing methods for DSA applications within the cognitive radio framework have been proposed in recent times [5]. These methods rely on the analysis and detection of spectral usage by primary users (PUs) in order to detect spectrum holes for utilization and include methods like matched filter sensing, cyclostationary feature based sensing, wavelet detection, compressed sensing [6], etc. However, these methods face several challenges which include: prior knowledge of receiver (modulation type, packet format, pulse shape etc.), primary carrier synchronization, increased hardware complexity (high resolution ADC, high speed processors etc.) as well as high power requirement ultimately leading to a significantly high cost of implementation. In such scenarios, the use of energy detection (ED) method [1], due to its simplistic approach, low computational cost and minimal hardware requirement is highly desirable. Moreover, since it does not rely on prior knowledge of the PU's signal, it is ideal for use in unknown radio environments. However, the efficacy of standard energy detection schemes which rely on a fixed threshold, is limited in conditions of low signal strength and high noise variance [7]. Channel conditions also play an important role in determining the sensing accuracy. This can be countered by use of an adaptive energy detection scheme which modifies the threshold by tracking variations in the active radio environment.

Many of these schemes have been proposed in literature and rely on parameters like spectrum width [1], signal interference and noise ratio (SINR), spectral efficiency and weight factor [8], etc. In this paper, we analyze and evaluate the performance of a SNR estimation based algorithm for adaptive and smart energy detection for spectrum sensing. Experiments are performed by designing the required system processes and models on the GNU radio and testing on the Universal Software Radio Peripheral (USRP B200) [9]. The SNR estimation technique gives better and reliable results than conventional energy detection techniques in case of low SNR conditions, thus improving overall performance of the system. The hardware complexity for this technique is also similar to conventional energy detectors, thus only a slight variation improves the performance of the overall CR system.

The remainder of the paper discusses about the several aspects of the system and is organized as follows: Section II talks about the system model and approach used for SNR estimation based energy detection algorithm, Section III gives the system description in terms of its hardware and architectural framework, Section IV discusses about the experimental results obtained after simulation and implementation of the system on USRP and finally, Section V concludes the paper.

## II. SYSTEM MODEL

For any spectrum sensing algorithm, the performance is measured in terms of both its sensitivity as well as its selectivity. Hence it is essential to take into consideration all the tradeoffs when setting the threshold for the energy detection scheme [10]. Let us quantify the received signal $y(n)$, a random variable, in the form of a binary hypothesis as shown below [8]:

$$y(n) = \begin{cases} w(n), & H_0 \\ s(n) + w(n), & H_1 \end{cases} \quad (1)$$

with $i = 1,2,3,...,N$, where $s(n)$ is the primary user signal while $w(n)$ is the AWGN noise signal with mean $\mu = 0$ and variance $\sigma^2$. Both the hypotheses represent the dual state of the channel on the basis of the value of the received signal, i.e. $H_0$ indicates that the primary user is present while $H_1$ indicates the absence. Thus the overall performance of such a cognitive radio system can be quantified by defining the values of $P_d$, i.e. the probability of successfully detecting the primary user as well as $P_f$, i.e. the probability of identifying a user when it does not exist in reality also known as a false alarm. Hence, a higher value of $P_d$ will ensure that there is minimum interference between primary and secondary users whereas a lower value of $P_f$ will ensure that there is good spectrum utilization.

For a generalized energy detector [11], the energy threshold for the received signal, $\lambda_{ge}$ is evaluated according to the following equation:

$$\lambda_{ge} = \frac{1}{N} \sum_{n=1}^{N} |y(n)|^p \quad (2)$$

where $y(n)$ is the received signal and $p$ is an arbitrary positive constant. This system becomes the conventional energy detector at $p = 2$. For this scheme, the spectrum sensing algorithm will evaluate the values of $P_d$ and $P_f$ as follows:

$$P_d = P_r\left(\lambda_{ge} < y(n) | H_1\right)$$
$$P_f = P_r\left(\lambda_{ge} < y(n) | H_0\right) \quad (3)$$

The ultimate goal of our spectrum sensing system is to increase the value of $P_d$ as well as reduce the value of $P_f$. However, this system is not adaptive in the sense that it does not modify its system characteristics like $\lambda_{ge}$ or $p$ based on the degradation of the active radio channel. In a condition of high noise variance as well as decreased SNR, this method is bound to fail. Hence, we adopt an adaptive approach to update the value of $\lambda_{ge}$ by constantly monitoring and estimating the value of SNR at various instants of time.

The adaptive threshold algorithm proposed in [12] is highly advantageous because it not only takes the SNR of the received signal into consideration but also estimates the noise variance and strives to provide a good value of $P_d$ even in conditions of dynamic noise variance. If we consider an AWGN channel, then the probabilities $P_f$ and $P_d$ based on the adaptive algorithm are given as follows [12]:

$$P_d = Q(\lambda - \mu_0/\sigma_0)$$
$$P_f = Q(\lambda - \mu_1/\sigma_1) \quad (4)$$

where Q(.) denotes the complementary error function, $\mu_0 = \sigma_n^2$, $\sigma_0 = \sigma_n^2/\sqrt{N}$, $\mu_1^2 = \sigma_0^2(\gamma + 1)$ and $\sigma_1 = \sigma_0^2(\sqrt{2\gamma + 1})$, where $\gamma$ denotes the given SNR at that instant. The upper and lower bounds of the variance $\sigma_n$ are also dependent on the the uncertainty factor $\rho$ as follows:

$$\sigma_l = \sigma_n * 10^{\frac{-\rho}{10}}$$
$$\sigma_h = \sigma_n * 10^{\frac{\rho}{10}} \quad (5)$$

Several SNR estimation strategies have been discussed in [13]. Accordingly, the maximum likelihood estimation technique for SNR evaluation is found to be ideal for the proposed sensing algorithm and implementation for the experiment. Based on the estimated value of the SNR, the threshold value $\lambda$ is iteratively updated by varying the number of samples accordingly [12]. Thus the final algorithm calculates the optimal number of samples required to successfully achieve the desired $P_d$ and $P_f$ for a given SNR value. Based on the above results we can easily evaluate the final value of the updated threshold energy which is shown below:

$$\lambda = \sigma_0 * Q^{-1}(P_f) + \mu_1 \quad (6)$$

Thus, the value of the threshold in the above equation changes based on the changing radio conditions. Fig. 1 shows the process flow for the above discussed iterative threshold adaption algorithm based on the estimation of the received SNR.

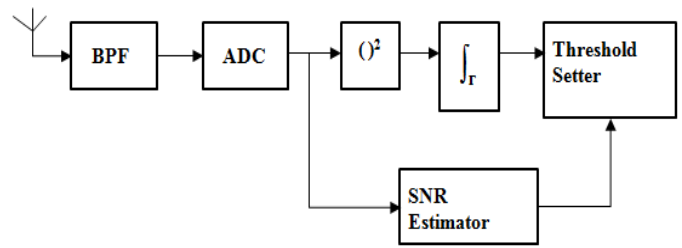

Fig. 1. System process for SNR estimation based adaptive energy detection.

The main advantage of this algorithm is that it is both channel independent as well as computationally inexpensive. The overall algorithmic flow for the proposed spectrum sensing method is shown in Fig. 2.

| **Algorithm:** Algorithm for SNR estimation based energy detection |
|---|
| **Input:** Estimated SNR |
| **Output:** Set adaptive threshold for radio environment |
| 1:   Estimate channel SNR using maximum likelihood estimation technique. |
| 2:   Compute $\mu_i$ as well as $\sigma_i$ for i=0 and i=1. |
| 3:   Set the desired value of $P_d$ and $P_f$. |
| 4:   Compute the number of samples required to achieve the value in Step 3. |
| 5:   Set the threshold energy $\lambda$ based on values evaluated in Step 1, Step 2, and Step 3. |
| 7:   Compute current channel statistics $y(n)$. |
| 8:      **if** $(y(n) > \lambda)$ **then** |
| 9:         Primary user detected. |
| 10:     **else if** $(y(n) < \lambda)$ **then** |
| 11:        Spectrum hole detected. |
| 12:     **end if** |
| 13:   Reinitialize SNR. |
| 14:   **return** spectrum attribute. |

Fig. 2. Overall algorithm for the spectrum sensing CR system.

## III. SYSTEM ARCHITECTURE

In this section, we discuss about the overall description as well as the architectural framework of our system. The implementation of the proposed SNR estimation based energy detection algorithm has been done on the USRP B200 which has been interfaced with GNU radio. Further details about individual blocks are given below.

### A. Universal Software Radio Peripheral

The USRP model B200 based GNU radio, shown in Fig. 3 [9] has been used to implement the above algorithm. Host connectivity has been established via a USB 3.0 controller and the UHD open source driver. MATLAB/SIMULINK connectivity has also been established to run a few simulations via the USRP support communication toolbox. Since the communication device can scan all the frequency bands that are FCC specified, the experiments were performed within the frequency band of 2.37 to 2.47 GHz as well as 0 to 5.7 MHz. Energy detection based on SNR estimation was used to identify spectrum holes in Wi-Fi as well as Bluetooth ranges.

### B. GNU Radio

The GNU Radio is an open source software that creates software defined radio (SDR) and signal processing systems using the signal processing blocks in its library [14]. It can work in two ways, first with an external hardware to create software defined radio and second, without any hardware to create a simulation like environment. C++ is used as the language to write the signal processing blocks. Python as a scripting language connects the blocks. The GNU Radio library consists of many readymade blocks. However, blocks can be created in the GNU Radio by the user to get desired results. The compilation of C++ and python languages is accomplished by using the SWIG interface compiler. The structure of GNU Radio and USRP SDR is shown in Fig. 4 [15]. As shown in the figure, the GNU Radio is connected to the USRP kit via a USB interface. USRP performs the function of an external hardware and provides input to the GNU Radio. Fig. 5 shows the overall system architecture as designed and modelled on GNU Radio.

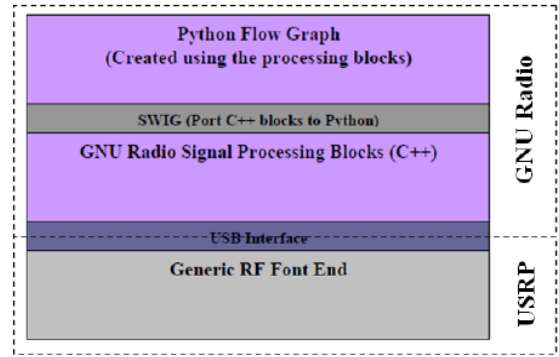

Fig. 4. Interlink between USRP and GNU Radio [15].

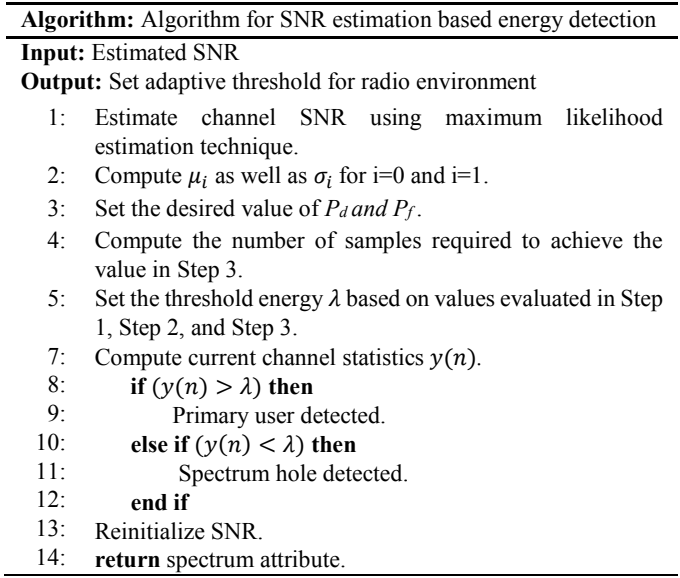

Fig. 3. The USRP B200 model.

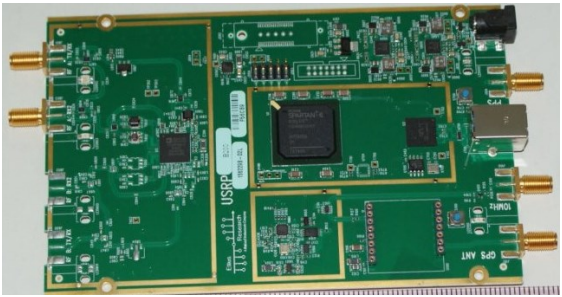

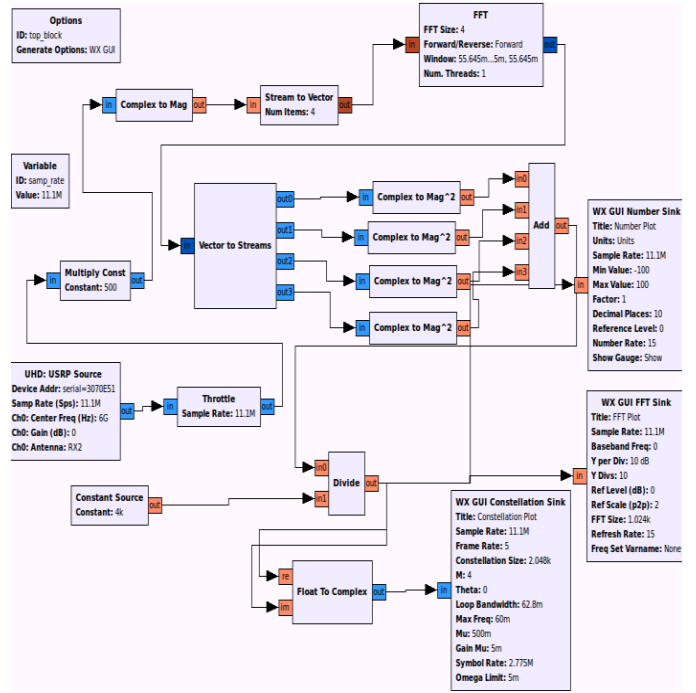

Fig. 5. Overall system modelling architecture of the SNR estimation based energy detection system for Cognitive Radio on GNU Radio and SDR.

IV. EXPERIMENTAL RESULTS

In this section, we analyze the overall performance of the spectrum sensing algorithm on the implemented hardware. The results were obtained using both conventional energy detection techniques as well as the SNR estimation based adaptive energy detection techniques. Both simulation as well as testing was done to monitor the overall system improvement in case of the new SNR based approach.

Fig. 6 compares the simulation results obtained for the ROC curves for both the general energy detection algorithm mentioned in equation (2) as well as the SNR estimation based adaptive threshold algorithm in equation (6). For similar conditions, the conventional model gives a maximum probability of detection of 0.2 while the SNR estimation method gives a probability of more than 0.8. We also observe that the adaptive threshold method is more reliable, especially in low SNR conditions, where the earlier method fails.

Since SNR estimation algorithm calculates the optimal number of samples required to achieve the desired value of detection rate or $P_d$, the number of samples required to achieve a fixed probability of detection of 0.7 was evaluated for various varying values of SNR. Fig. 8 shows the graph between the

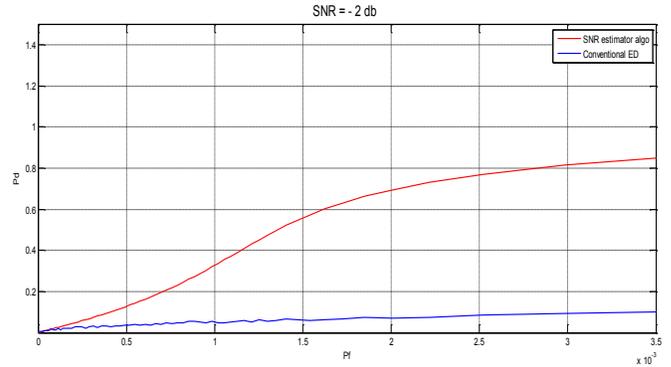

Fig. 6. ROC curve of $P_d$ vs $P_f$ for adaptive SNR algorithm (red) as well as conventional energy detection algorithm (blue).

number of samples M vs the SNR. It can be observed, that it is easier to arrive at a decision metric when the SNR is high, and hence the number of samples required to be evaluated is significantly less as compared to when the SNR is very low. In case of conventional energy detection schemes, the number of samples required for accurate prediction stays constant irrespective of the dynamic changes in the radio environment. This leads to inefficient spectral usage as the same CR network will draw constant power irrespective of the noise variance. However, the use of SNR based adaptive threshold technique allows for the dynamic allocation of computing resources, thus

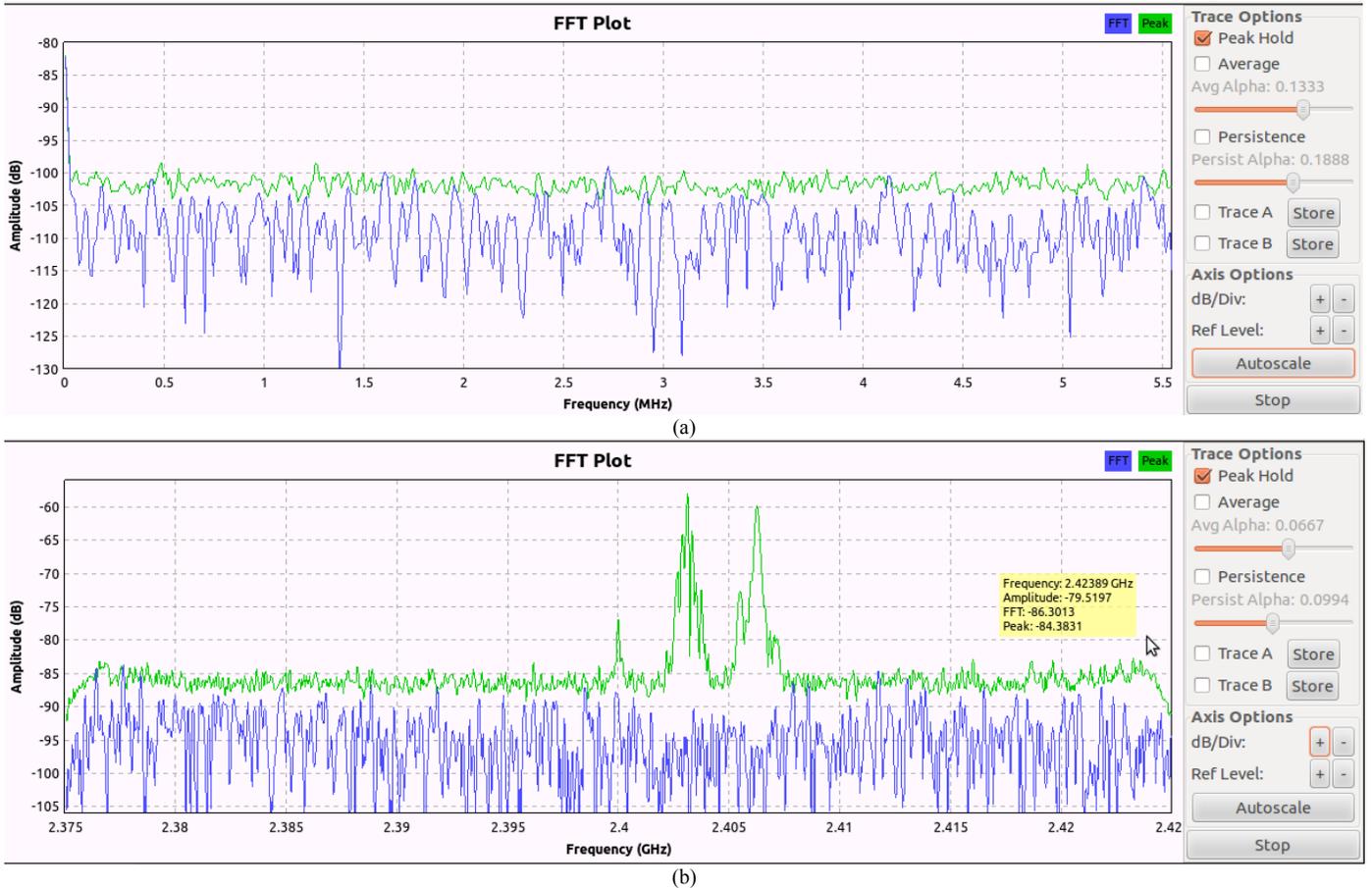

Fig. 7. The FFT plot for real time USRP sensing in: (a) The 0-5.5 MHz band using the conventional energy detection algorithm, (b) The 2.3-2.4 GHz band using adaptive SNR estimation based algorithm. Notice the two peaks around the 2.405 GHz frequency in case of the adaptive SNR estimation based algorithm.

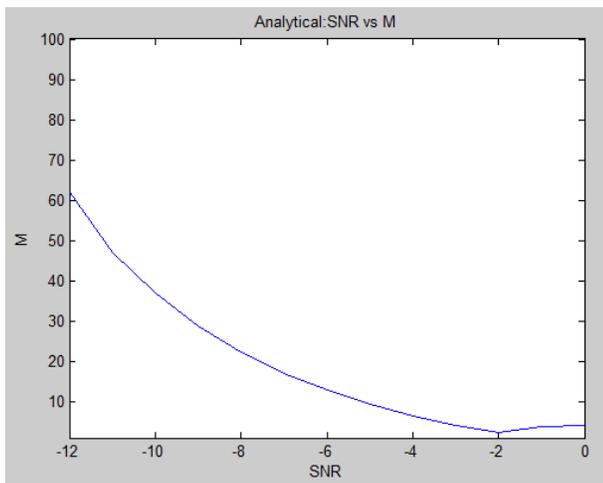

Fig. 8. Graph of number of samples M vs the SNR value.

working in sync with the radio environment leading to better system efficiency.

Real time implementation of both conventional as well as the adaptive models was done on two frequency bands respectively. The USRP along with the GNU radio was used to evaluate and plot the FFT response of the received signal strength over the specified bands. Fig. 7 (a) shows the result of the conventional energy detection technique for spectrum sensing in the 0 to 5.5 MHz band. The result shows that no specific peak is detected by the transceiver when using the standard technique. The high variance of noise, in blue, along with low SNR of the signal contributed to the lack of detection of any peak within the spectrum. This, in a real life scenario would lead to strong interference with the primary user and would reduce the final quality of service. Fig. 7 (b) however, shows the same result for the SNR estimation based energy detection scheme used for the Wi-Fi spectral band of 2.3 to 2.4 GHz. We can observe specific spectral use at 2.402 GHz as well as 2.407 GHz respectively. Thus, we also identified the spectral window available for secondary use at 2.405 GHz. The use of SNR estimation based energy detection improves the overall performance of the CR system by many folds and is highly reliable in case of low SNR or high noise variance. Thus, the adaptive technique to iteratively update the threshold value for energy detection gives satisfactory results in a wide variety of environments as compared to the conventional energy detection technique which fails upon channel degradation.

## V. CONCLUSION

Only a slight variation in the conventional energy detection technique improves the spectrum sensing performance and efficiency of the cognitive radio by many folds. The use of SNR estimation based energy detection has improved the performance of the cognitive radio in low SNR and high noise variance conditions. The proposed method was implemented on the USRP B200 and GNU radio giving improved performance. While the standard energy detection method failed to register any peaks, the SNR estimation based algorithm registered two peaks as well as located a spectral window in the 2.4 GHz band. The adaptive approach has the benefit of not only increased system performance but also at reduced receiver complexity. Thus the proposed system has a wide scope of development in the implementation and use in a variety of cognitive radio infrastructures. Future scope for development exists in design and development of smarter CR frameworks, which focus not only on cognitive sensing but smart reconfigurability as well.

ACKNOWLEDGMENTS

The authors are indebted to Prof. Khyati Vachhani for her valuable guidance and inspiration in the successful completion of this work.